\documentclass[fleqn,twoside]{article}
\usepackage{espcrc2}

\newcommand{\AmS}{{\protect\the\textfont2
  A\kern-.1667em\lower.5ex\hbox{M}\kern-.125emS}}
  
\newcommand{\lap}{\lower.5ex\hbox{$\; \buildrel < \over \sim \;$}}
\newcommand{\gap}{\lower.5ex\hbox{$\; \buildrel > \over \sim \;$}}

\title{Open Problems in Cosmology}

\author{P. J. E. Peebles\address[?]{Joseph Henry Laboratories,  Princeton University, Princeton NJ 08544 USA}}

\begin{document}

\begin{abstract}
The main task for cosmology today is the completion of the program of cosmological tests that commenced in the 1930s. I offer comments on the state of the program and the issues to be addressed to establish a convincing large-scale picture of the observable universe.
\vspace{1pc}
\end{abstract}

\maketitle

\section{INTRODUCTION}

Many cosmologists today are fully occupied sorting out the implications of our new standard Friedmann-Lema\^\i tre $\Lambda$CDM cosmology. There is good reason: this cosmology has passed demanding tests. Cosmology still is an attempt to draw very large conclusions from exceedingly limited data, however, and it is proper that searches for departures from the standard cosmology continue. But the intensity of the debates is notably subdued compared to decades past, a result of the great observational advances and their general agreement with a remarkably simple model.  

Many of the ideas in the standard model are notably durable. Einstein's empirical guides to general relativity theory came from the length scales of the laboratory, in the E\"otv\"os experiment and electromagnetism. Now his theory passes demanding tests on the length scales of cosmology. Lema\^\i tre introduced the idea that that the universe expanded from a Big Bang -- his primeval atom --  in 1931. Now we have convincing evidence this is what actually happened. Hubble's book,  {\it The Realm of the Nebulae} (1936), shows a cosmological test, a galaxy redshift-magnitude diagram ($z$-$m$ for short, where the cosmological redshift $z$ is defined by the ratio of observed to emitted wavelengths, $1+z=\lambda _{\rm obs}/\lambda _{\rm em}$, and the apparent magnitude is $m = -2.5\log _{10}f + {\rm constant}$, where $f$ is the observed energy flux density). Now an enormous effort with vastly more capable technology applied to observations of supernovae has increased the limiting redshift of the $z$-$m$ test by a factor of ten. Tolman might be gratified to learn that his (1934) theory of the behavior of thermal electromagnetic radiation in an expanding homogeneous and isotropic universe is now applied to precision measurements. In the 1930s Dirac considered evolution of the strength of the gravitational interaction. Models and tests for variations of the strengths of the gravitational and electromagnetic interactions were much discussed in the 1960s, led largely by Jordan and Dicke, in attempts to give meaning to Mach's principle and to serve as a foil to general relativity in the new generation of tests of gravity physics. Now the idea is driven by superstring scenarios.  There are unexpected developments, to be sure. I have seen no evidence Tolman gave thought to the effect of the departures from a homogeneous mass distribution on the distribution of the thermal radiation. Now this is a key cosmological test. Many people have given thought to the cosmological constant, $\Lambda$, since its introduction in 1917, often with negative conclusions. Now we have significant evidence for its detection. I doubt Einstein or Lema\^\i tre or Dirac argued for weakly interacting dark matter. Now we have strong evidence for its existence, though we await substantiation by a laboratory detection and identification. 

It is also worth bearing in mind that observational advances have ruled out elegant ideas; the steady state cosmology is an example. And for all we know other ideas have survived only because we have not yet found out what is wrong with them: longevity adds a patina of respectability. In short, we should respect the ideas behind the standard cosmology, because elegant ideas sometimes lead to aspects of reality, but we should use caution in adding ideas to the established cannon, because the universe is quite capable of surprising us. My comments on what might be done to improve the cannon are organized around a list of issues to be addressed. 

\section{CHALLENGES}

The peremptory style of the headers in this list should not be misunderstood: people are working on these challenges and need no exhortations from me to carry on. But they might find it helpful to reflect on what people are doing about associated problems, and the rest of us may find the headers a useful overview of issues that stand in the way of a satisfactory completion of our understanding of the large-scale nature of the observable universe.

\subsection{Systematize tests of gravity physics}

Superstring theory has inspired thoughts of departures from Einstein's general relativity theory (GR for short), and that has led to experimental tests. But the  more general consideration is that it is an enormous extrapolation from the characteristic scales of the precision tests of GR, $\lap 10^{13}$~cm, to the application on the scales of cosmology, $\sim 10^{28}$~cm. Good science demands checks; this is a purpose of the cosmological tests.

The present most convincing application of the cosmological tests follows from the measurements of the large-scale mean mass density.  We have abundant cross checks -- many of which are discussed in other contributions to these Proceedings -- from gravitational dynamics in systems ranging in size from the Local Group to the Local Supercluster of galaxies; gravitational lensing; the cluster baryon mass fraction; the large-scale galaxy correlation function; the number density of rich clusters of galaxies as a function of time and mass;  the SNeIa redshift-magnitude ($z$-$m$) relation; the intermediate Sachs-Wolfe effect; and the ratio $H_ot_o$ of stellar evolution and expansion time scales. The results consistently indicate the mean mass density is
\begin{equation}
\Omega _m = 0.25\pm 0.1,
\end{equation}
expressed as a fraction of the mass density at critical escape velocity. 

The error flag is large: this is not precision cosmology. But we have a good case that it is accurate cosmology, that is, a good approximation to physical reality. This is because the different methods depend on quite different aspects of the theory to convert the observations into constraints on $\Omega _m$, and a wrong theory is not likely to mislead us so systematically, from so many different directions. For example, the first entry in the list -- the gravitational dynamics of relative motions of gas and stars and galaxies -- depends on the assumption that starlight is a useful tracer of mass. That seems reasonable, because one gets consistent results from a considerable range of length scales, but it certainly may be questioned (and generally was a few years ago, when the Einstein-de Sitter model with $\Omega _m=1$ was popular). The next entry, gravitational lensing, depends less heavily on starlight as a tracer of mass, and the consistent constraint on $\Omega _m$ is a valuable check. Both assume the inverse square law for gravity, with the relativistic factor of two for lensing. We have checks, from the supernova $z$-$m$ relation, and the ratio $H_ot_o$ of stellar evolution and expansion time scales, both of which indicate the mass density is in the range in equation~(1). Both depend on GR,  but applied to the evolution of the spatially homogeneous cosmological model, quite a different aspect of the physics from gravitational dynamics and lensing. Other entries in the list depend on the cold dark matter (CDM) theory for structure formation, so the model must be added to the theory to be tested. We have tests, from the theory and observation of the anisotropy of the thermal 3~K background radiation, the abundance of rich clusters as a function of mass and redshift, and the large-scale galaxy correlation function.  One could change the last constraint by tilting the spectrum of large-scale primeval mass density fluctuations away from the standard CDM model, but that would mess up the successful fit to the 3~K anisotropy measurements.

The measurements also depend on interpretations of the astronomy, which is chancy. Remember the astronomer's Tantalus effect: you may look but never touch. Thus $H_ot_o$ depends on the extragalactic distance scale, which is very difficult to establish, and the supernova $z$-$m$ relation depends on inferences about the properties of stars that exploded when the world was young, not an easy task. But errors in interpreting the astronomy are not likely to so systematically mislead us. 

I offer two conclusions. 

\noindent 1. The abundant cross checks offer compelling evidence that the parameter $\Omega _m$ is meaningful and its value constrained to a factor of two or so. This goes along with the conclusion that GR and the CDM are useful approximations to this aspect of reality. 

\noindent 2. It is time to systematize the tests of GR, maybe following the example of the parametrized post-Newtonian formalism used in tests of gravity physics on the scales of stellar systems. The PPN formalism commences with Minkowskian spacetime, and uses parametrized representations of the effects on spacetime of its material content. In cosmology might start from the Robertson-Walker line element, which follows from the observed symmetry, and introduce parametrized relations between the material content of the universe and its expansion history and departures from the homogeneous line element. One would explore how closely the observations constrain the parameters, and whether the constraints contain the GR values. 

\subsection{Tighten the evidence for $\Lambda$}

As just discussed, a dense web of evidence establishes the case that the measure of the mean mass density in equation~(1) is meaningful. The case for detection of Einstein's cosmological constant $\Lambda$ (or a term in the stress-energy tensor that acts like it) is significant too, but not as thoroughly checked. The wonderfully successful measurement and theoretical interpretation of the anisotropy of the 3~K thermal radiation, which Carlstrom discusses in these Proceedings, does not yet strongly constrain $\Omega _m$, but it does require small space curvature. Since we have abundant evidence that $\Omega _m$ is less than unity, GR requires a cosmological constant (or its equivalent) with $\Omega _\Lambda = 1 - \Omega _m$. We have a check, from the supernova $z$-$m$ relation, which favors a low density cosmologically flat universe with $\Omega _\Lambda = 1 - \Omega _m$ over a low density open universe with $\Omega _\Lambda =0$. 

It takes nothing away from the magnificent accomplishments of the supernova observers to say that their present measurements alone would not make a strong case for detection of $\Lambda$: the astronomy is too dicey. The evidence from the 3~K anisotropy is strong but not compelling, in  my opinion, because I don't know how we can be sure there is no variant of the CDM theory -- maybe with delayed recombination, maybe evolving coupling parameters -- that fits the measurements in a low density open universe. The idea seems unlikely, but so does $\Lambda$.  

The merits of the cosmological constant have been debated for a long time, and we can wait a few more years to see how the web of evidence develops, perhaps from extensions of the measurements of the supernova $z$-$m$ relation to higher redshift, perhaps from an examination of the difference between the intermediate Sachs-Wolfe effect in flat and open low density universes, perhaps from improvements in the measurement of  $H_ot_o$. Meanwhile we owe our support -- from funding to TACs -- to colleagues who are working so hard on this key issue. 

\subsection{Find the physics of the vacuum energy} 

The challenge was known in the 1930s. Pauli is said to have remarked that the zero-point energy of the electromagnetic field integrated over laboratory wavelengths would, in the static Einstein model, require curvature that does ``not even reach to the Moon.'' Since then particle physics has given us a lot more contributions to the zero-point energy, along with the latent heats of all the first-order phase transitions. 

I have heard suggestions that the problem may lie in the imperfect method of computation of energy in curved and dynamical spacetime, on the right hand side of Einstein's field equation, or maybe even with the left hand side, which says how the stress-energy tensor drives the geometry. This certainly must be considered, but it looks like a long shot to me. I think of the quantum theory of electromagnetism as standard textbook quantum mechanics. We have firm experimental evidence that zero-point energies contribute to the active gravitational mass of matter, and I don't see why the same would not be true of the zero-point energy of the electromagnetic field. And we know the large-scale distribution of ordinary matter and radiation affects the expansion history of the universe about as predicted by GR.  But the implication from this reasoning is absurd. 

I hear discussions of two remedies. Maybe some symmetry principle forces the vacuum energy density to the only natural and acceptable value, zero, and the  astronomers' $\Lambda$  has nothing to do with vacuum expectation values. Or maybe the net vacuum energy density, represented by $\Lambda$, is a slowly varying function of position, and we flourish where we can, near a place where the vacuum energy density passes through zero, just as we flourish on the one planet in the Solar System where complex life as we know it is possible. I hear complaints that this anthropic principle has been introduced {\it ad hoc}, to save the phenomenon. But the same is true of $\Lambda$. The cosmological constant is now seen to save quite a few phenomena. And a justification of the anthropic principle would be convincing  evidence that the vacuum energy density actually is a function of position. Something may turn up. 

\subsection{Sort out the cosmic coincidences}

We have to live with some curious coincidences in the standard cosmology. Most widely noted nowadays is that we flourish just as the universe is making the transition away from matter-dominated expansion -- in order of magnitude. Under the anthropic principle this is not so curious, of course, because our presence does not require that $\Lambda$ vanish, only that its absolute value is small enough to allow galaxies to form and harbor the several generations of supernovae that produce the chemicals life needs. I am not inclined to count this as evidence for the anthropic principle. But have you heard a better idea?

Another coincidence of time scales is that we flourish just as the Milky Way is running out of gas for the formation of new stars and planetary systems. Maybe this is related to the observation that the global star formation rate has begun rapidly decreasing. Maybe both are related to the transition away from matter-dominated expansion, which suppresses the gravitational accretion of extragalactic matter to supply material for new generations of stars. But this does not seem promising in the $\Lambda$CDM model, where the suppression effect is quite modest. I have not heard discussions of the stronger suppression in a low density universe with $\Lambda =0$. 

Yet another time coincidence is the agreement of the age of the expanding universe with the classical electron radius multiplied by the ratio of the electric to gravitational forces between an electron and proton. Dirac pointed this out in the 1930s. Still another, in the $\Lambda$CDM model for structure formation, is that galaxies have just become useful tracers of  mass, with the added curiosity that the low order galaxy n-point correlation functions seem to be better approximations to simple power laws than are the predicted correlation functions of the mass that is supposed to be controlling the dynamics. 

It's reasonable to expect that some of these curiosities are purely accidental, and that some will be seen not to be curious at all when we really understand the physics and its relation to the observations. But it's sensible to be aware that some might be clues to improvements in the physics. 

\subsection{Test the physics of the dark sector}

The physics of particles and fields in the visible sector is wonderfully simple, its expression spectacularly complex. In the standard cosmology the physics of the dark sector -- which includes 96\%\ of the mass --  is so  simple that its expression is simple: the dark matter is a collisionless gas that gathers into slightly lumpy dark halos, and $\Lambda$ or its equivalent is constant or close to it. Is this the way it is, or only an useful approximation for the purpose of our crude observational constraints?

The particle physicists' assignment is to learn whether the dark matter is detectable. Its identification within an enlarged  standard model for particle physics could settle the question. The astronomers' assignment is to check whether the standard model for the dark sector gives a full and complete account of the observations of structure formation and evolution. The astronomers' problem is that structure formation in the visible sector is complicated, so they have to interpose a model between the physics of the dark sector and its expression in the observations. This means they must discover whether anomalies in the theory and observation of structure formation are only apparent, from flaws in the model, and whether real anomalies might be hidden by the parameter adjustments allowed by the model. Dealing with this is a rich challenge, but it will be met by the rich fund of empirical evidence astronomers are accumulating on how the galaxies formed.

It is good strategy for astronomers to operate under the assumption that the dark sector is well described by simple physics unless or until anomalies point to deficiencies and, with luck, hints to remedies. But it is good strategy also to anticipate the possible need for remedies, by exploring adjustments to the physics of the dark sector that are viable within the observational constraints. You can find examples on arXiv; it makes sense to be aware of this line of work. 

\subsection{Find the physics of high redshift}

What was the physics of the very early universe, before it could have been described by the relativistic Friedmann-Lema\^\i tre cosmology? Ideas about the answer inform assessments of the present structure of the universe. 

The curious nature of initial conditions for the Friedmann-Lema\^\i tre model -- the relativistic singularity and the disconnect between large-scale homogeneity and the small particle horizon at high redshift -- was known well before inflation, but not widely advertised, in large part because we didn't know what to do about it. It became famous with the introduction of a plausible resolution, the inflation concept, in the early 1980s, which you might have thought would have stimulated thinking along other lines, but it was not until the early 1990s that we started hearing about scenarios inspired by superstring theory. Will we be hearing about still more lines of thought? How might we decide which if any is a useful approximation to what really happened in the very early universe? 

Pre-inflation speculations about bouncing or oscillating universes sometimes assumed the bounce conserves baryons but maybe not entropy. Recent ideas take the opposite tack: physics outside the Friedmann-Lema\^\i tre model produced entropy and the entropy produced the baryons, with negligibly small neutrino chemical potential and a homogeneous ratio of the local baryon number to photon number densities. A demonstration from established particle physics that this is the correct version of baryogenesis would be beneficial in fixing ideas in cosmology. Detection of the scale-invariant spectrum of gravitational waves predicted by some inflation models -- from the effect on the CBR polarization, and maybe by means of a gravitational wave detector in space -- would be an influential argument for inflation. Periodic universes have been under discussion for a long time, and remain an active subject for research. Detection would be startling and deeply influential. Nowadays you don't hear much about the idea that we stand to learn something about a Big Bang and Big Crunch from the physics and astronomy of the little crunches in stellar mass black holes and active galactic nuclei, but fashons may change again. Modern discussions of structure formation in cosmology usually accept as established the initial conditions of the $\Lambda$CDM cosmology. What about the cosmic strings and monopoles and textures that were so cried up a decade ago? These isocurvature departures from homogeneity are not likely to have been the primary initial conditions for structure formation, but they certainly could play a secondary role, and maybe lead us to a better cosmology. 

The subtext is the hope that something may turn up to add to our meager store of empirical hints to the physics of the early universe. I see a  parallel to the situation in the 1930s that is reflected in Hubble's concluding statement in {\it The Realm of the Nebulae}: ``The search will continue. Not until the empirical resources are exhausted, need we pass on to the dreamy realms of speculation.'' We are much further down the road, but in the same situation. Like Hubble, we don't know whether the empirical resources are close to being exhausted, but we can be confident the search will continue.
 
\end{document}